\documentclass[11pt,a4paper]{article}
\pdfoutput=1
\usepackage{graphicx}
\usepackage{setspace}
\usepackage{amsmath,amssymb,color,mathrsfs,verbatim,bbm,wasysym,pstricks,epsfig,slashed}
\usepackage{jheppub}
\usepackage{graphicx}
\usepackage{caption}
\usepackage{subcaption}

\newcommand{\reliso}{\mathcal{R}_\textrm{iso}^\ell}

\title{Identifying boosted new physics with non-isolated leptons}
\author[a,b,c]{Christopher Brust,}
\author[a]{Petar Maksimovic,}
\author[a]{Alice Sady,}
\author[a,b]{Prashant Saraswat,}
\author[a,d]{\\Matthew T.\ Walters}
\author[a]{and Yongjie Xin}
\affiliation[a]{Department of Physics and Astronomy, Johns Hopkins University, \\
Charles Street, Baltimore, MD 21218, U.S.A.}
\affiliation[b]{Department of Physics, University of Maryland, \\
Campus Drive, College Park, MD 20742, U.S.A.}
\affiliation[c]{Perimeter Institute for Theoretical Physics, \\
Caroline Street N, Waterloo, Ontario, N2L 2Y5, Canada}
\affiliation[d]{Department of Physics, Boston University, \\
Commonwealth Avenue, Boston, MA 02215, U.S.A.}
\emailAdd{cbrust@perimeterinstitute.ca, petar@pha.jhu.edu, asady1@jhu.edu, saraswat@umd.edu, mtwalter@bu.edu, yongjie@pha.jhu.edu}
\abstract{We demonstrate the utility of leptons which fail standard isolation criteria in searches for new physics at the LHC. Such leptons can arise in any event containing a highly boosted particle which decays to both leptons and quarks. We begin by considering multiple extensions to the Standard Model which primarily lead to events with non-isolated leptons and are therefore missed by current search strategies. We emphasize the failure of standard isolation variables to adequately discriminate between signal and SM background for any value of the isolation cuts. We then introduce a new approach which makes use of jet substructure techniques to distinguish a broad range of signals from QCD events. We proceed with a simulated, proof-of-principle search for $R$-parity violating supersymmetry to demonstrate both the experimental reach possible with the use of non-isolated leptons and the utility of new substructure variables over existing techniques.}
\begin{document}

\begin{flushright}UMD-PP-014-011\end{flushright}

\maketitle
\flushbottom

%%%%%%%%%%%%%%%%%%%%%%%%%%%%%%%%%%%%%%%%%%%%%%%%%%%%%%%%%%%%%%%%%%%%%%%%%%%%%%%%
%%%%%%%%%%%%%%%%%%%%%%%%%%%%%%%%%%%%%%%%%%%%%%%%%%%%%%%%%%%%%%%%%%%%%%%%%%%%%%%%
%%%%%%%%%%%%%%%%%%%%%%%%%%%%%%%%%%%%%%%%%%%%%%%%%%%%%%%%%%%%%%%%%%%%%%%%%%%%%%%%

\section{Introduction}
\label{sec:intro}

One of the primary goals of the CMS and ATLAS experiments is to search for physics beyond the Standard Model~\cite{CMS:1994hea,Armstrong:1994it}. To achieve this it is advantageous to use every possible handle on new physics that could help distinguish it from Standard Model (SM) processes. In this paper we discuss a potential new physics signature which is currently underutilized by LHC searches, namely non-isolated leptons. These are (charged) leptons which are produced at small angular separation from hadronic activity, including hard jets. In physics beyond the Standard Model (BSM), this can occur preferentially if the lepton is produced along with colored partons from the decay of a particle with high Lorentz boost. Some examples are models with heavy resonances decaying to top quarks ($Z' \rightarrow t\bar{t}$), in which the top quarks are produced at high boost and can decay leptonically, or supersymmetric models with light, highly boosted neutralinos which can decay through $R$-parity violation to leptons and quarks ($\tilde{\chi}_0 \rightarrow q q \ell$).

However, non-isolated leptons are also a common feature of Standard Model QCD jets, as some hadrons in a jet can produce leptons in their decays. This occurs particularly often in jets originating from heavy quarks. % such as bottoms ($b$'s).
In order to cut down on this QCD background, most current LHC searches (e.g.~\cite{Chatrchyan:2012ira, Aad:2014qaa, CMS:2014qpa, TheATLAScollaboration:2013cia} and many others) only use leptons that are isolated from hadronic activity for purposes of event selection. While this is appropriate when targeting leptons originating from e.g.\ unboosted $W$ or $Z$ bosons, this requirement can preclude the selection of new physics events with leptons from boosted objects. Current leptonic searches can therefore have very little sensitivity to such models. New physics of this type could in principle be present in the 8 TeV LHC data, without being identified by existing searches (as demonstrated for RPV SUSY models in~\cite{Graham:2014vya}).

In this work we discuss how to achieve sensitivity to new physics producing non-isolated leptons. To this end we examine the lepton isolation criteria currently in use by the ATLAS and CMS experiments and consider alternative approaches. Our main result is a proposal for a new substructure-based variable, lepton subjet fraction (LSF), to distinguish leptons arising in decays of boosted new physics from those produced in QCD jets. This new variable, combined with appropriate hard kinematic cuts, allows one to reject most of the QCD background while maintaining high acceptance for signal events.

In section \ref{sec:motivation}, we discuss the generic form of models which tend to produce non-isolated leptons and present multiple well-motivated examples. In section \ref{sec:variables}, we discuss the limitations of standard isolation variables and introduce our new substructure discriminant for removing standard backgrounds. We then use these variables to compare the dominant SM backgrounds to various example models of new physics. In section \ref{sec:search}, we consider a proof-of-principle search for one of these models (the $\tilde{q}\rightarrow q \tilde{\chi}_0$ model discussed in \cite{Graham:2014vya}) to show the effectiveness of our new strategy. In section \ref{sec:conclusions}, we conclude with some discussion of the implementation of this strategy at future runs of the LHC, as well as possible future directions for study.

%%%%%%%%%%%%%%%%%%%%%%%%%%%%%%%%%%%%%%%%%%%%%%%%%%%%%%%%%%%%%%%%%%%%%%%%%%%%%%%%
%%%%%%%%%%%%%%%%%%%%%%%%%%%%%%%%%%%%%%%%%%%%%%%%%%%%%%%%%%%%%%%%%%%%%%%%%%%%%%%%
%%%%%%%%%%%%%%%%%%%%%%%%%%%%%%%%%%%%%%%%%%%%%%%%%%%%%%%%%%%%%%%%%%%%%%%%%%%%%%%%

\section{Theoretical motivation}
\label{sec:motivation}

At hadron colliders such as the LHC, selecting for leptons in the final state is a powerful approach to distinguish signal events from QCD background. Current searches for new physics by ATLAS and CMS generally require these leptons to be isolated as parameterized by the relative isolation variable:
\begin{equation}
\reliso = \frac{\sum_i p_{T,i}}{p_{T,\ell}},
\label{eq:RelIso}
\end{equation}
where $i$ sums over all non-leptonic particles within a cone of some radius $R_{\textrm{cone}}$ of the lepton $\ell$. Typically, the isolation requirement is $\reliso \lesssim 0.1 - 0.2$ with cone radius $R_{\textrm{cone}} \sim 0.3 - 0.4$ (e.g.~\cite{Chatrchyan:2012ira, Aad:2014qaa, CMS:2014qpa, TheATLAScollaboration:2013cia}). Such a requirement is very effective in rejecting leptons arising from decays of hadrons in QCD jets, while retaining high acceptance for most types of new physics signals. However, there are classes of BSM models in which hard processes can produce leptons that are typically non-isolated (high $\reliso$) and are therefore poorly constrained by current searches.  

%The purpose of this isolation requirement is to distinguish leptons produced in the decay of a new particle from those arising in the formation of jets. Unsurprisingly, this discriminant is quite effective at identifying signal events for many models, significantly reducing the dominant SM backgrounds. However, there are many well-motivated new physics scenarios which generically produce non-isolated leptons. The purpose of this work is to argue that such models are discoverable at both the 8 and 13 TeV runs of the LHC, but are {\it currently being missed} by standard analyses.

In such models, signal events feature production of a heavy new particle $X$ that decays to some much lighter particle $L$, which in turn decays to colored partons and at least one lepton. Since $L$ is produced at high boost in the lab frame, the leptons will be kinematically restricted to align with the jet(s) from the colored partons, such that they fail standard isolation requirements. There are many possibilities for $X$ and $L$ as well as their decay products; below, we consider three distinct scenarios to demonstrate the broad range of theories containing such spectra. This set of examples is intended to be motivational rather than exhaustive. 

%For momentum conservation, $X$ must also decay to one or more additional particles $Y$. The full chain of decays is therefore $X \rightarrow L + Y$, $L \rightarrow \ell + nj$. 

\begin{figure}[t!] 
\centering
\includegraphics[width=8.0cm]{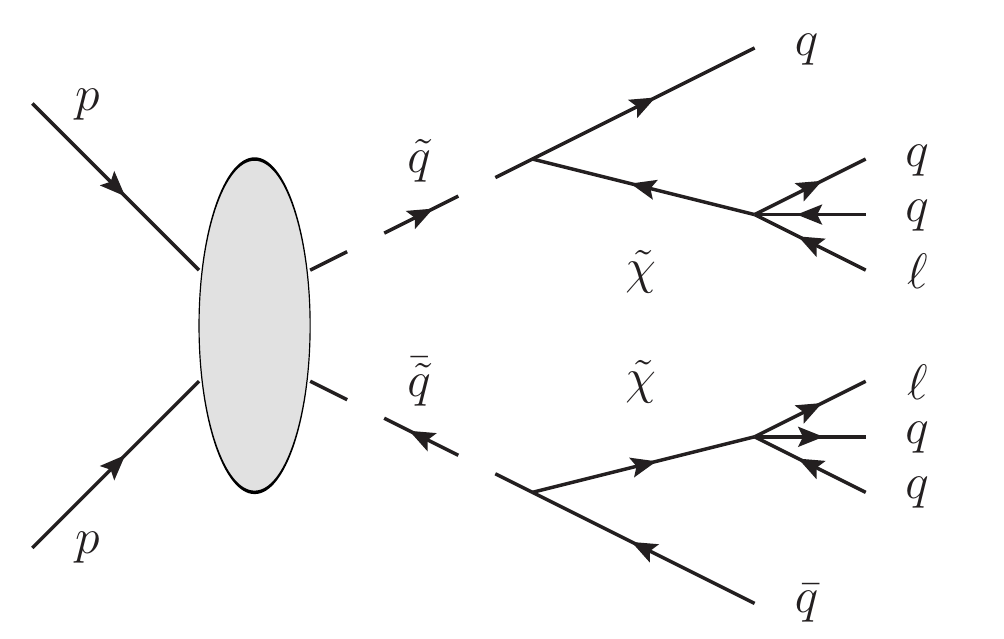}
\caption{Signal topology for the squark-neutralino ($\tilde{q}\rightarrow q \tilde{\chi}$) model. This model yields non-isolated leptons when the squark-neutralino mass splitting is large enough for the neutralino to be significantly boosted.}
\label{fig:squark-topology}
\end{figure}

One possibility is for $L$ to be a new, undiscovered light particle. If this particle has weak couplings to SM fields, then its direct production cross-section can be low enough to evade current searches even when its mass is relatively low.  The production cross-section of the heavy $X$ particle however may be much larger (e.g.\ if it is colored), such that the $L$ is predominantly produced at high boost. Such phenomenology can occur in supersymmetric (SUSY) models, in which a weakly coupled neutralino can be produced at high boost from decays of much heavier squarks. With $R$-parity violation (RPV), the neutralino can decay through an off-shell slepton or squark to $\ell qq$ via the lepton-number-violating $LQD$ operator, as shown in figure \ref{fig:squark-topology}, so that SUSY events become rich in non-isolated leptons~\cite{Graham:2014vya}. In section \ref{sec:search}, we present a simulated search for this particular model at the 8 TeV LHC, demonstrating the greatly increased signal sensitivity possible with the use of non-isolated leptons. 

%(Note that if $L$ is an $SU(3)$ singlet, then it must decay to more than one colored parton.)

%However, its direct production cross-section must be low enough to evade current bounds from searches for isolated leptons. We therefore need the dominant production mechanism for $L$ to be through decays of $X$, such that $L$ can only be produced with significant boost.

%The simplest way to suppress direct production of $L$ is for it to be a singlet under $SU(3)_C$. A simple example from SUSY is a neutralino, which can be produced from the decay of a much heavier squark. In a simplified model with R-parity violation (RPV), the neutralino can then decay through an off-shell slepton or squark to $qq \ell$ with the $QLD$ operator, as shown in figure \ref{fig:squark-topology}. Note that in such a scenario the full set of decay products must be color-neutral, generically requiring two or more jets in the final state.

\begin{figure}[t!] 
\centering
\includegraphics[width=8.0cm]{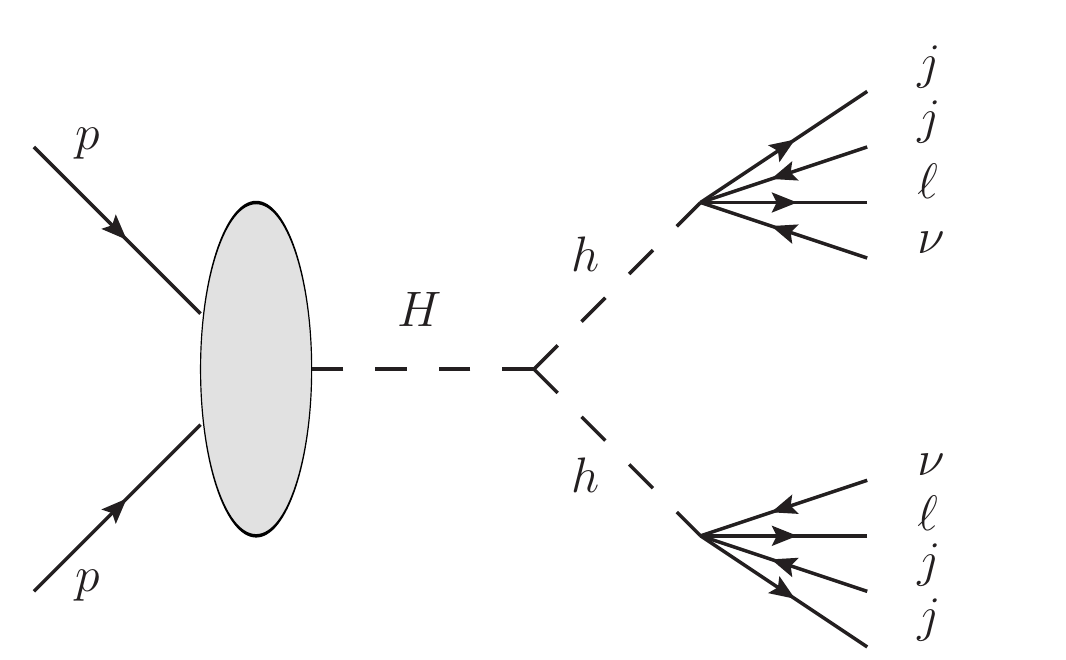}
\caption{Signal topology for the $H\rightarrow hh$ model. This model yields non-isolated leptons when the $H$ mass is high enough for the Higgs pair to be significantly boosted.}
\label{fig:diHiggs-topology}
\end{figure}

A separate possibility is for $L$ to be a SM particle which can decay to leptons and quarks, such as the Higgs ($h \to WW^* \to qq\ell\nu$) or top ($t \to bW \to b\ell\nu$). One potential example of such a model is a heavy scalar $H$ which decays to a pair of Higgs bosons, as shown in figure \ref{fig:diHiggs-topology}. Such heavy scalars generically arise within the set of ``two Higgs doublet'' models~\cite{Lee:1973iz,Branco:2011iw}, leading to non-isolated leptons in the boosted Higgs decays. Another example in SUSY models is ``stoponium'', stop-antistop bound states that can form if the lightest stop is stable, and which can annihilate to Higgses \cite{Barger:1988sp, Kumar:2014bca}. Similarly, boosted tops are potential signatures of new resonances which dominantly couple to heavy quarks, such as Kaluza-Klein gluons or electroweak gauge bosons in models with warped extra dimensions~\cite{Agashe:2006hk, Lillie:2007yh,Agashe:2008jb}, or new gauge bosons in ``topcolor'' models~\cite{Chivukula:1998wd}.  As shown in previous work, the use of non-isolated leptons to identify these boosted tops can significantly improve the signal efficiency and reduce the background, extending the potential discovery reach~\cite{Thaler:2008ju,Rehermann:2010vq}. Note that, unlike the other scenarios we have discussed, in boosted top models the leptons are produced in association with a single quark rather than two, which is only possible because the top itself carries color. 

%Another well-motivated possibility is a massive vector boson $Z'$, arising from a new broken $U(1)$, which predominantly decays to the top quark

In the following section, we compare these potential signals to the dominant SM backgrounds which produce non-isolated leptons. While we limit our discussion to these specific examples, we expect that our results are representative of a broader class of accessible models, implying that non-isolated leptons can be a useful general probe of physics beyond the SM. %We therefore hope the use of non-isolated leptons yields a promising avenue of research for discovering new physics.

%%%%%%%%%%%%%%%%%%%%%%%%%%%%%%%%%%%%%%%%%%%%%%%%%%%%%%%%%%%%%%%%%%%%%%%%%%%%%%%%
%%%%%%%%%%%%%%%%%%%%%%%%%%%%%%%%%%%%%%%%%%%%%%%%%%%%%%%%%%%%%%%%%%%%%%%%%%%%%%%%
%%%%%%%%%%%%%%%%%%%%%%%%%%%%%%%%%%%%%%%%%%%%%%%%%%%%%%%%%%%%%%%%%%%%%%%%%%%%%%%%

\section{Lepton isolation variables}
\label{sec:variables}

We now consider the general problem of discriminating between leptons arising from boosted signal objects and those produced in QCD jets. Although specific searches will often need to place hard kinematic cuts on events, in this section we aim to demonstrate the degree of signal versus background discrimination that can be achieved using lepton isolation variables alone. To this end, we consider the distributions of these variables for signal and background events after relatively loose kinematic cuts, namely requiring one lepton with $p_T >$ 20 GeV (for both 8 and 13 TeV events) and two jets with $p_T >$ 150 GeV (8 TeV) or $p_T >$ 250 GeV (13 TeV).

For these distributions, we initially place no isolation requirement, such that any lepton with $p_T > 20$ GeV and $|\eta| < 2.5$ will not be clustered into a jet. Henceforth, we shall only use the term ``lepton'' to refer to an electron or muon satisfying these kinematic cuts, with no restrictions on isolation. Anything that fails these baseline cuts is then rolled into the hadronic activity of the event, which is clustered into jets using the anti-$k_T$ algorithm~\cite{Cacciari:2008gp} with radius parameter $R_\textrm{jet}=0.5$, as well as the minimal jet requirements $p_T > 20$ GeV and $|\eta| < 3.0$. The events were generated using MadGraph 5.1.5.7 \cite{Alwall:2011uj}, showered and hadronized with Pythia 8.1.85 \cite{Sjostrand:2007gs}, and clustered using Fastjet 3.0.6 \cite{Cacciari:2011ma}. Our analysis is performed at ``truth level'' for an individual $pp$ collision, without simulation of detector uncertainties or pileup effects.

%We wish to identify leptons buried in a large amount of hadronic activity. Typically, such leptons would fail a relative isolation requirement and be clustered into the surrounding jets. We shall consider a different approach. Rather than separate leptons into isolated and non-isolated, we will have {\it no} isolation requirement, such that {\it any} lepton with $p_\textrm{T} > 20$ GeV and $|\eta| < 2.5$ will not be clustered into a jet. Conversely, anything that fails these baseline cuts is then rolled into the hadronic activity of the event. Henceforth, we shall only use the term ``lepton'' to refer to an electron or muon with $p_\textrm{T} > 20$ GeV and $|\eta| < 2.5$, with no restrictions on isolation.

\begin{figure}[t!] 
\centering
\includegraphics[width=7.0cm]{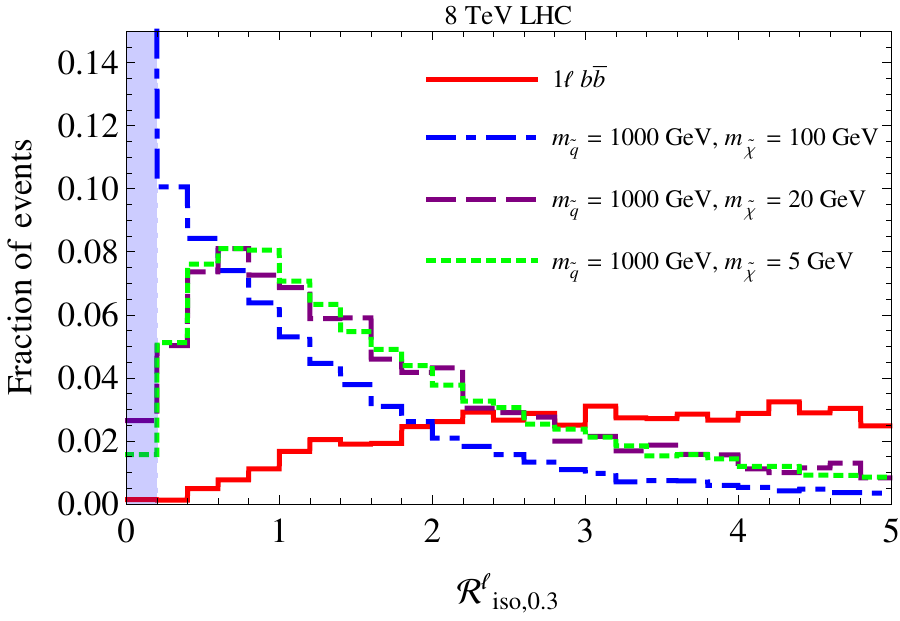}
\hspace{1.5mm}
\includegraphics[width=7.0cm]{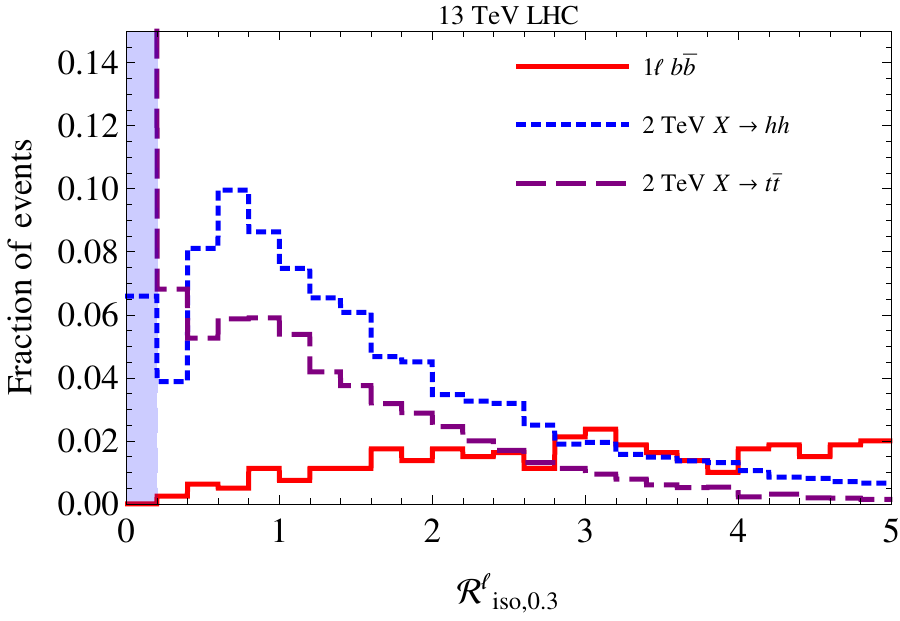}

\vspace{5 mm}

\includegraphics[width=7.0cm]{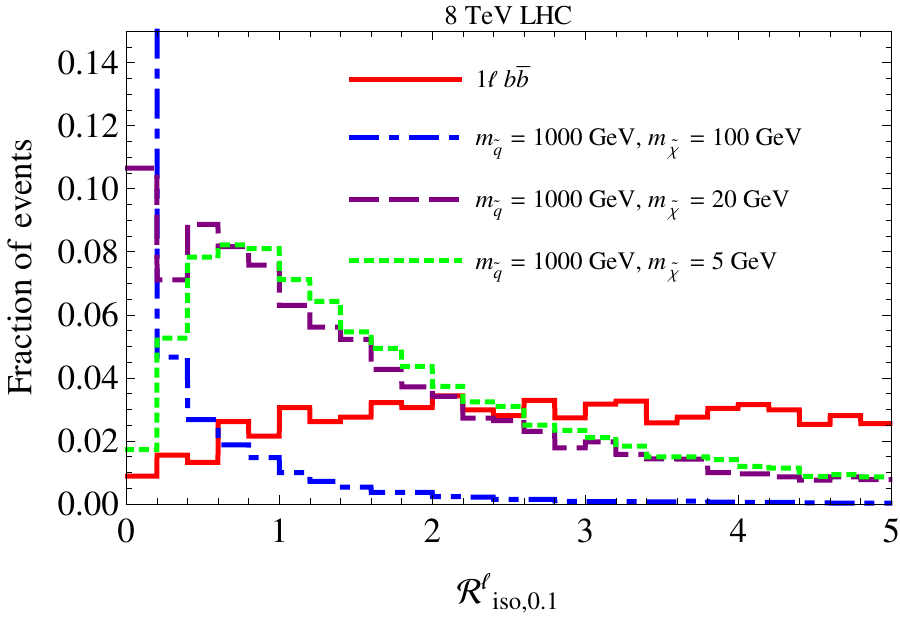}
\hspace{1.5mm}
\includegraphics[width=7.0cm]{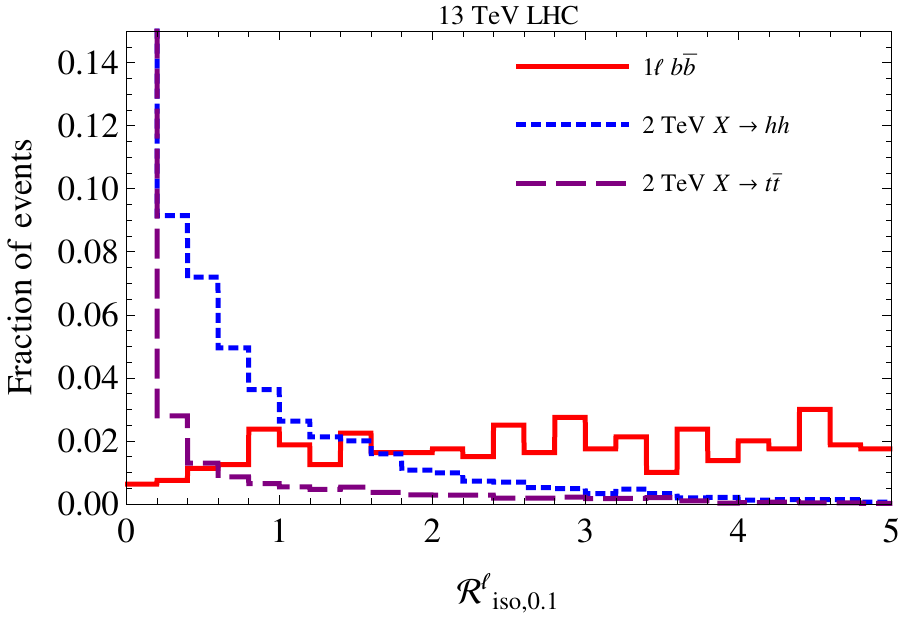}
\caption{Unit-normalized distributions of $\reliso$ of the leading lepton for various event samples. In the upper plots, $\reliso$ is computed with a cone size $R_{\textrm{cone}} = 0.3$, while in the lower plots $R_{\textrm{cone}} = 0.1$. The left plots show distributions at 8 TeV for the $b\bar{b}$ sample (solid red) and the $\tilde{q}\rightarrow q \tilde{\chi}$ models with $m_{\tilde{q}} = 1000$ GeV and $m_{\tilde{\chi}} = 100$ GeV, 20 GeV, and 5 GeV. The right plots show distributions at 13 TeV for $b\bar{b}$ and for models of a 2 TeV resonance decaying to Higgses (dotted blue) or tops (dashed purple). The shaded blue regions indicate the fraction of events captured with the typical isolation cut $\reliso<0.2$ with $R_{\textrm{cone}}=0.3$.}
\label{fig:RelIso}
\end{figure}

%To overcome this one may consider computing $\reliso$ with smaller cone sizes. Figure~\ref{fig:RelIso1} shows distributions of $\reliso$ as in figure~\ref{fig:RelIso4} but with a cone size $R$ = .1. For moderately boosted signals we see that signal efficiency remains very high even with tight cuts on this variable. However the highly boosted signals still have limited discrimination from background.       

%The standard variable used by ATLAS and CMS to define lepton isolation is the relative isolation parameter $\reliso$ discussed above (eq.\ \ref{eq:RelIso}).
Figure \ref{fig:RelIso} shows the distribution of the standard relative isolation parameter $\reliso$, defined with cone sizes of $R_\textrm{cone} = 0.3$ (upper plots) and $R_\textrm{cone} = 0.1$ (lower plots), for various signal and background samples. As we are interested in distinguishing signal events from QCD jets containing leptons, we specifically consider the background due to $b\bar{b}$ production, which often gives a hard non-isolated lepton. An example of a typical cut used in LHC searches (e.g.~\cite{Chatrchyan:2012yca}) is $\reliso < 0.2$ with $R_\textrm{cone} = 0.3$, corresponding to the blue shaded region on the upper plots in figure~\ref{fig:RelIso}. While this cut overwhelmingly rejects the $b \bar{b}$ background, it also removes most of the signal, particularly for the highly boosted squark-neutralino model. For the moderately boosted signals we see that signal efficiency can be recovered while still retaining good background discrimination by considering looser cuts on $\reliso$ and/or computing $\reliso$ with a smaller cone of size $R_\textrm{cone} = 0.1$. However, the highly boosted signals (here, the squark-neutralino models with neutralino masses of 20 GeV and 5 GeV) are difficult to distinguish from background with either variation of this variable. 

%\footnote{We again cluster {\it only} those leptons that fail baseline cuts of $p_T > 20$ GeV, $|\eta| < 2.5$, never clustering any harder leptons into jets, regardless of isolation.}

\begin{figure}[t!] 
\centering
\includegraphics[width=7.0cm]{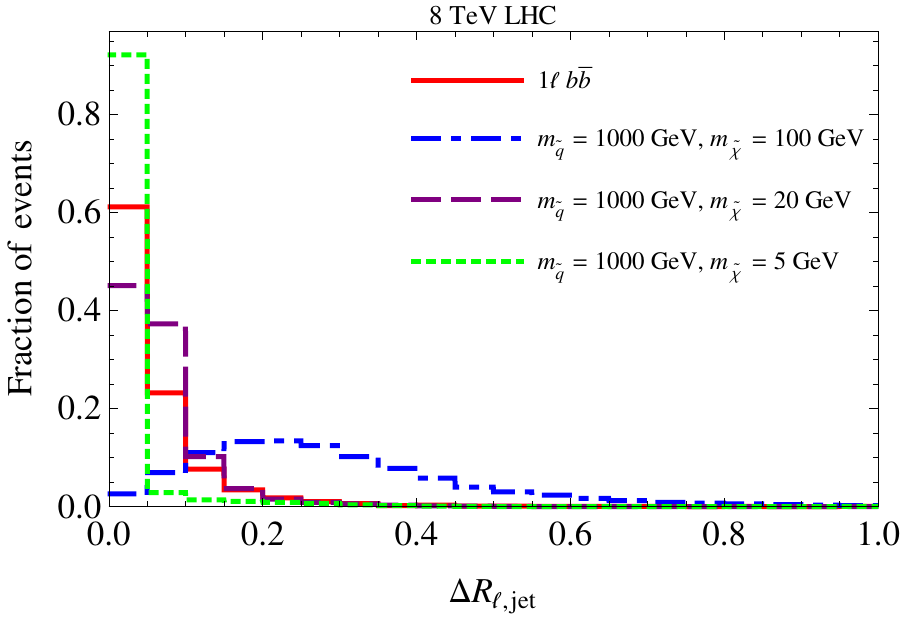}
\hspace{1.5mm}
\includegraphics[width=7.0cm]{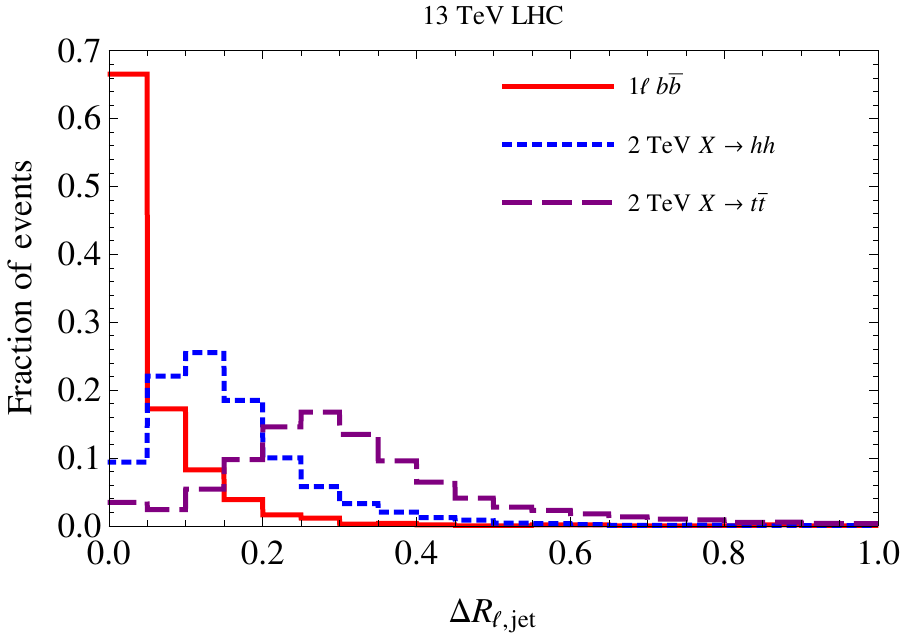}
\caption{Left: Distributions of $\Delta R_{\ell,\text{jet}}$ for the leading lepton at 8 TeV for the $b\bar{b}$ sample (solid red) and the $\tilde{q}\rightarrow q \tilde{\chi}$ models with $m_{\tilde{q}} = 1000$ GeV and $m_{\tilde{\chi}} = 100$ GeV, 20 GeV, and 5 GeV. Right: Distribution of the same variable at 13 TeV for $b\bar{b}$ and for models of a 2 TeV resonance decaying to Higgses (dotted blue) or tops (dashed purple).}
\label{fig:DeltaR}
\end{figure}

The inherent limitations of $\reliso$ for these signals can be understood by considering the angular separation $\Delta R_{\ell,\text{jet}}$ between the lepton and the closest jet axis, which has been used as a simple loosened isolation variable by some LHC searches~\cite{CMS:2009cxa,Chapleau:2010nn,Chatrchyan:2012cx,Aad:2013nca}, and has been proposed previously as one of several variables intended to discover non-isolated leptons in boosted leptonic tops \cite{Thaler:2008ju}. (As stated above, we do \emph{not} include hard leptons in the jet clustering itself.) Distributions of this variable are shown in figure~\ref{fig:DeltaR}. For moderately boosted signals, there is significant angular separation between the lepton and the jet axis, such that the lepton is relatively isolated. However, for the highly boosted signals, the lepton is in fact typically \emph{closer} to the jet axis than in $b \bar{b}$ events, unlike in boosted top events. This is to be expected, since in the limit of high boost the lepton is highly collimated with both of the original partons giving rise to the jet, and therefore must be aligned with the jet axis. In $b$-jets, however, leptons arise from the decay of a $b$-hadron, which although highly boosted is typically displaced from the axis of the jet by $\mathcal{O}(\alpha_s)$ due to showering. Therefore, in terms of this measure the leptons from very highly boosted signals are in fact less ``isolated'' than those from QCD jets.    

It is clear then that any lepton isolation criterion that requires some fixed angular separation of the lepton from hadronic activity will not accept sufficiently boosted signals. However, even in the case of high boost there is a physically different relationship between the lepton and the jet in signal events compared to QCD jets. In signal events, the lepton is produced in the hard process and has ``split off'' from the quarks before parton showering. The lepton is then a bystander to the subsequent evolution of the jet. In contrast, in QCD/heavy-flavor jets, leptons originate from hadrons or radiated photons; they are essentially products of the parton shower and are correlated with the pattern of radiation within the jet. This suggests that jet substructure information could be useful to differentiate signal and background jets, inspiring our subjet-based approach.

\begin{figure}[t!] 
\centering
\includegraphics[width=7.0cm]{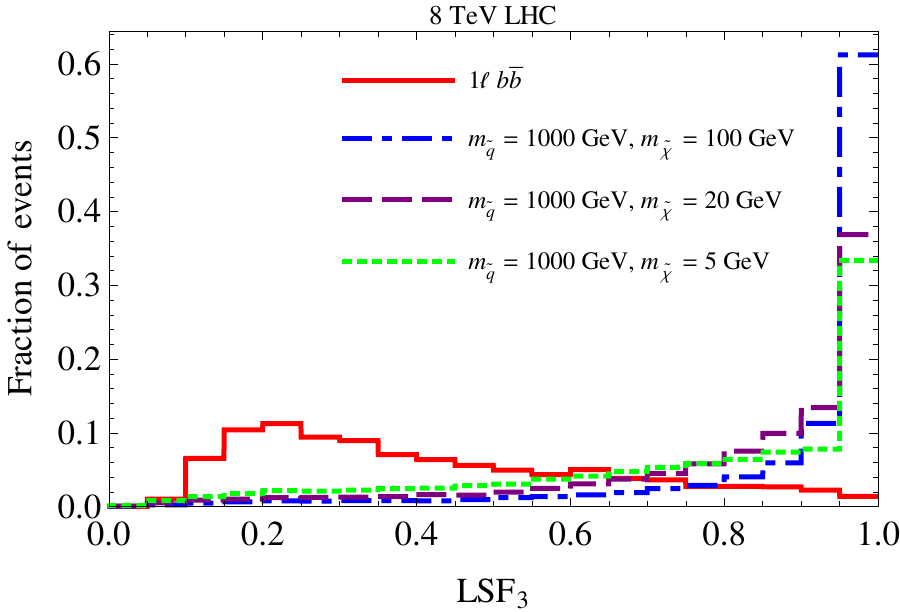}
\hspace{1.5mm}
\includegraphics[width=7.0cm]{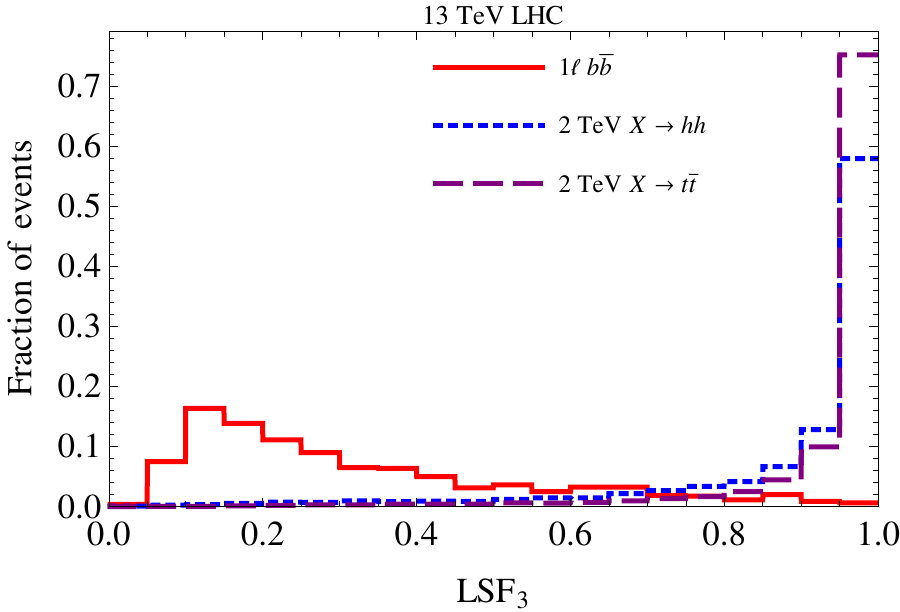}
\caption{Left: Distributions of $\text{LSF}_3$ for the leading lepton at 8 TeV for the $b\bar{b}$ sample (solid red) and the $\tilde{q}\rightarrow q \tilde{\chi}$ models with $m_{\tilde{q}} = 1000$ GeV and $m_{\tilde{\chi}} = 100$ GeV, 20 GeV, and 5 GeV. Right: Distribution of the same variable at 13 TeV for $b\bar{b}$ and for models of a 2 TeV resonance decaying to Higgses (dotted blue) or tops (dashed purple).}
\label{fig:LepRatio}
\end{figure}

Based on this observation, we propose a new lepton isolation variable: the lepton subjet fraction (LSF). This variable is computed by first clustering \emph{all} particles in an event, including all leptons, into ``fat jets'' using the Cambridge-Aachen algorithm~\cite{Dokshitzer:1997in} with radius parameter $R_\textrm{fat} = 0.8$. (This clustering can be considered completely independently of the clustering used to define signal jets for an analysis, and is then merely an intermediate step in computing LSF.) For each fat jet, we then cluster its constituents into $n$ subjets using the exclusive $k_T$ algorithm \cite{Catani:1993hr}, with $n$ optimally chosen to match the number of subjets expected in signal events. This algorithm proceeds by defining a distance measure $d_{ij} = \text{min}(p_{T,i}^2,p_{T,j}^2) \Delta R_{ij}$ between all pairs of particles, clustering the pair with minimum $d_{ij}$ into a single pseudoparticle, and iterating until only $n$ pseudoparticles are left, which are identified as the $n$ subjets. Once this is done, all leptons in the event have been associated with a subjet. The lepton subjet fraction of each lepton is defined as the ratio of the lepton $p_T$ to its associated subjet $p_T$:
\begin{equation}
\text{LSF} = \frac{p_{T,\ell}}{p_{T,sj}}.
\end{equation}      
Leptons with higher LSF are considered to be more isolated. We plot distributions of LSF for our various samples in figure~\ref{fig:LepRatio}, with $n = 3$ (henceforth we use the label $\text{LSF}_n$ to specify LSF as computed with a specific number of subjets $n$) . We see that even extremely boosted signals are quite differentiated from background in this variable, being peaked towards LSF of unity, while $b \bar{b}$ events have a mostly flat tail at high LSF.

The success of this variable can be understood by considering how $k_T$ clustering typically acts upon signal and background jets. The $k_T$ algorithm tends to cluster soft radiation first, eventually converging onto the hardest cores of energy in the jet. In signal events, the lepton is usually one of the hardest individual particles in the jet and therefore tends to be left alone until late stages of the clustering. Early on it will only be grouped with soft particles that accidentally happen to be closer to the lepton than to any other particles. The other hadronic particles then tend to cluster as they would in the absence of the lepton, essentially rewinding the parton shower back to the hardest splittings between colored partons. These hard splittings arise from both QCD emission as well as the original decay of the boosted parent to a lepton and quarks; for low boost the latter will be the largest splitting while for very high boost the former dominates. Either way, the lepton has no tendency to align with the products of these splittings, and thus usually dominates one of the final $n$ subjets produced by the clustering. In contrast, in background events the lepton tends to be aligned with one of the hardest cores of energy within the jet and ends up with more hadronic energy clustered with it, even when it is relatively energetic. As there is \emph{no} size parameter in exclusive $k_T$ clustering, this effective discrimination continues to hold even at high boost.

%Some additional insight can be gained by considering another variable,

Some additional information is contained in another variable, the lepton mass drop (LMD) of the subjet associated with a lepton, defined as\footnote{This ratio of invariant masses is based on the ``mass drop'' parameter introduced in \cite{Butterworth:2008iy}.} 
\begin{equation}
\text{LMD} = \frac{m_{sj-\ell}}{m_{sj}},
\end{equation}  
where $m_{sj}$ denotes the invariant mass of the subjet including the lepton and $m_{sj-\ell}$ denotes the mass of the subjet with the lepton subtracted out, i.e.\ the mass of the hadronic component only. In the limit that the latter mass is much less than the energy of the subjet with the lepton subtracted out, $\text{LMD}$ is kinematically required to be less than $1 - \text{LSF}$, with the limit achieved when the lepton is collinear with the total subjet. Note therefore that isolated or ``signal-like'' leptons are defined as those with \emph{low} LMD. 

\begin{figure}[!t] 
\centering
\includegraphics[width=7.0cm]{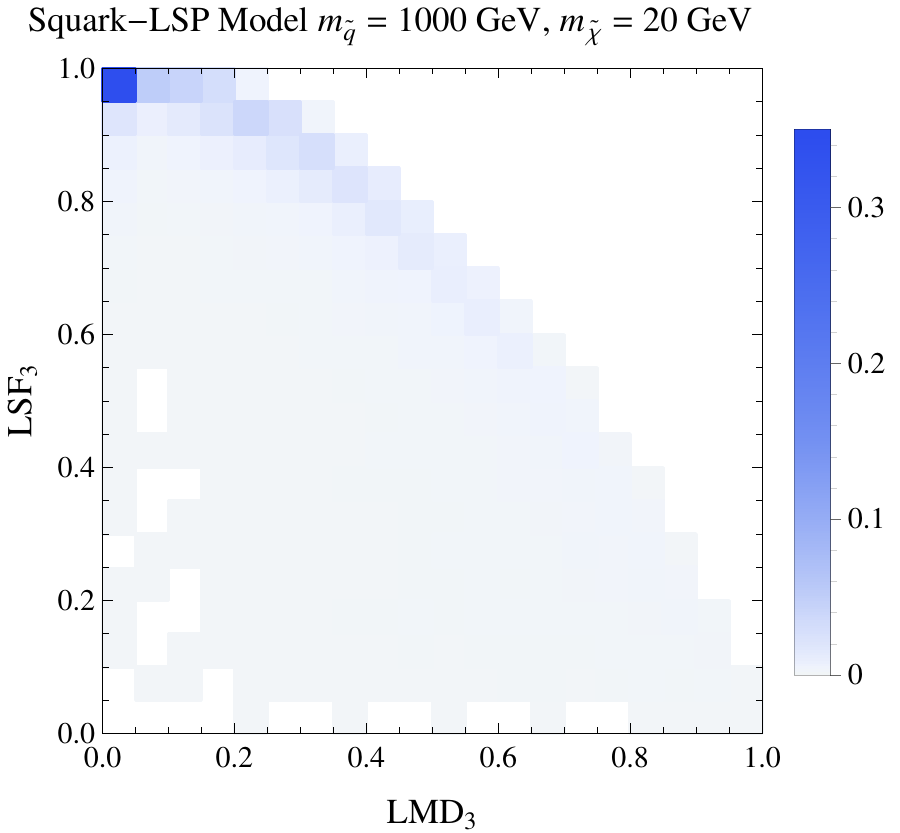}
\hspace{1.5mm}
\includegraphics[width=7.2cm]{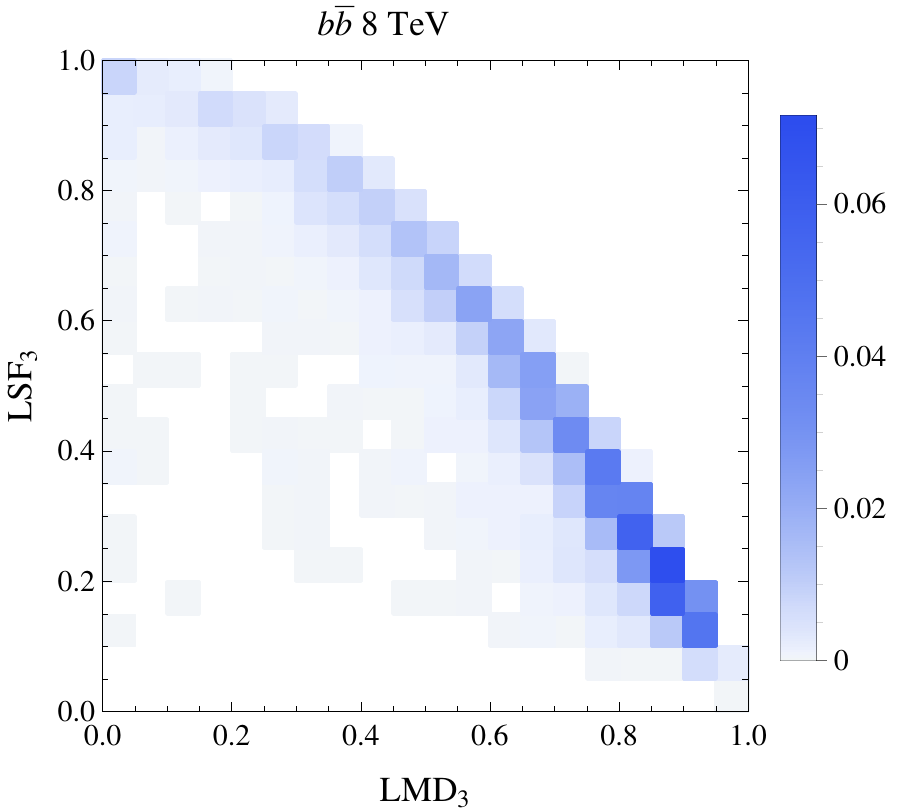}
\caption{Unit-normalized 2D histogram of $\text{LSF}_3$ and $\text{LMD}_3$ for the leading lepton at 8 TeV, for the $\tilde{q}\rightarrow q \tilde{\chi}$ model with $m_{\tilde{q}} = 1000$ GeV and $m_{\tilde{\chi}} = 20$ GeV (left) and for the $b\bar{b}$ sample (right).}
\label{fig:LepRatiovsMassRatio}
\end{figure}

In figure~\ref{fig:LepRatiovsMassRatio} we plot distributions in the plane of LSF and LMD for the squark-neutralino signal and the $b \bar{b}$ background, both at 8 TeV. In both cases we see that these two variables are highly correlated, tending to lie near the approximate kinematic limit, such that either could be used to define an isolation cut. Because of this correlation, at this level of analysis cutting on both LSF and LMD simultaneously would appear to give at most a modest gain in signal discrimination. However, this strategy may be useful for other models or when detector effects are accounted for. For example, in the presence of pileup a lepton is likely to be clustered with some stray hadronic activity even if it would otherwise be completely isolated; in such a case the lepton may have LSF less than unity yet have LMD very close to zero due to the large typical angle between the lepton and pileup activity.

In the previous discussion we focused on LSF as defined by clustering fat jets into three subjets ($\text{LSF}_3$), which is appropriate for the signal models we have considered in which boosted particles decay into up to three visible partons. The behavior of LSF defined with a different number of subjets $n$ is easily understood. Consider $\text{LSF}_2$, i.e.\ continuing the $k_T$ clustering for an additional step so that only two subjets remain. For signals with only two visible partons produced in the decay of the boosted object (such as boosted tops), the two final subjets tend to include one dominated by the lepton and one dominated by the products of the other, colored parton, so that the LSF is still close to unity (though the inequality $\text{LSF}_m \leq \text{LSF}_n$ for $ m < n$  always holds). However, for signals with three distinct partons from the boosted decay, clustering the fat jet into only two subjets has an $\mathcal{O}(1)$ chance of clustering the lepton subjet with a subjet arising from one of the other partons, greatly decreasing LSF and eroding the ability to discriminate signal. Similarly we may consider clustering into more than three subjets (e.g.\ $\text{LSF}_4$). In this case the LSF would remain near unity for both two-parton and three-parton signals. However, for background jets there is potential for LSF to increase significantly if more subjets are retained, so search sensitivity may be compromised.

\begin{figure}[!t] 
\centering
        \begin{subfigure}[b]{7.0cm}
                \includegraphics[width=7.0cm]{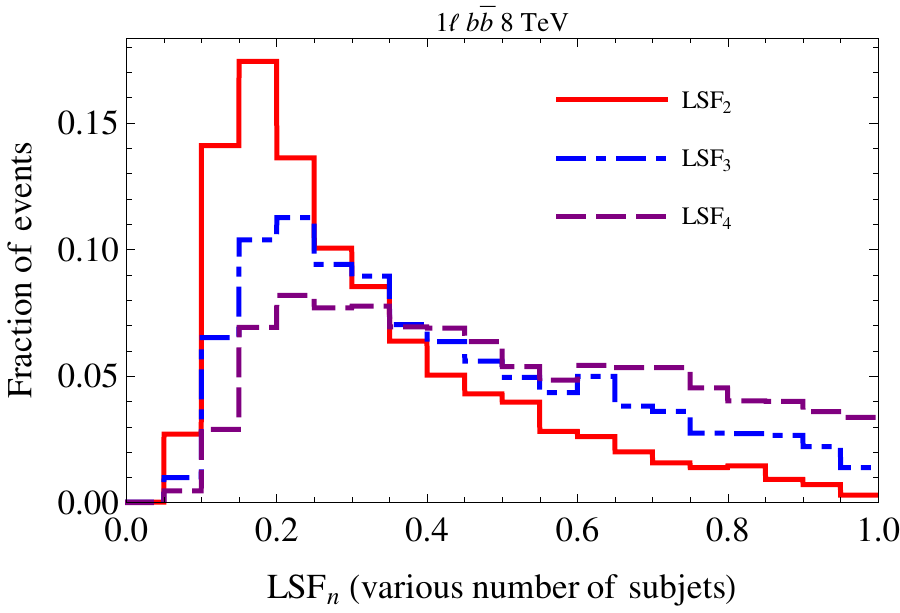}
                \vspace{-5mm}            
               \caption{}
                \label{fig:LSFn_bb}
        \end{subfigure}
        \begin{subfigure}[b]{7.0cm}
\hspace{5mm}
\includegraphics[width=7.0cm]{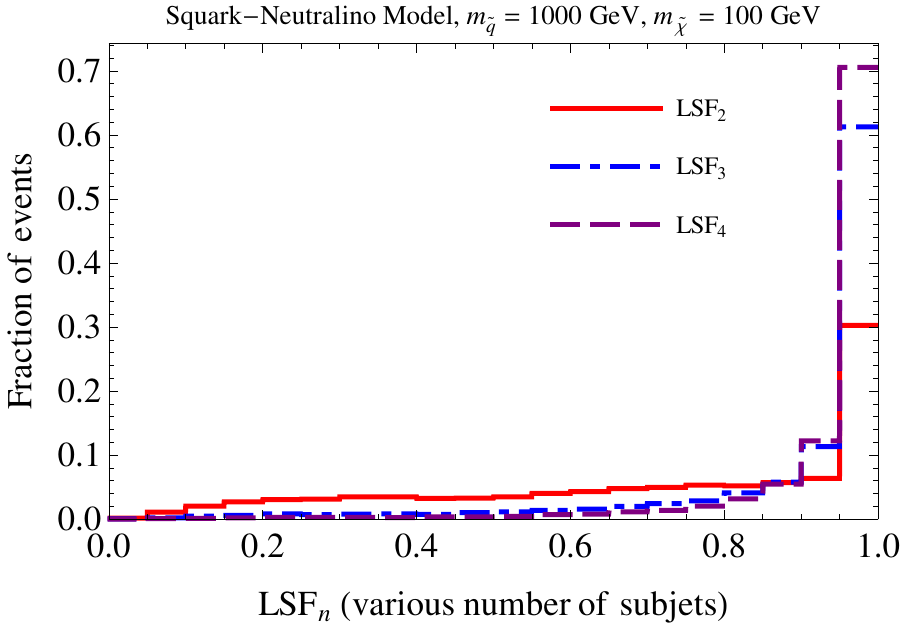}
                \vspace{-5mm}
                \label{fig:LSFn_squark}
                \caption{}
        \end{subfigure}

\vspace{5mm}

       \hspace{2.8mm}
        \begin{subfigure}[b]{7.0cm}\includegraphics[width=7.0cm]{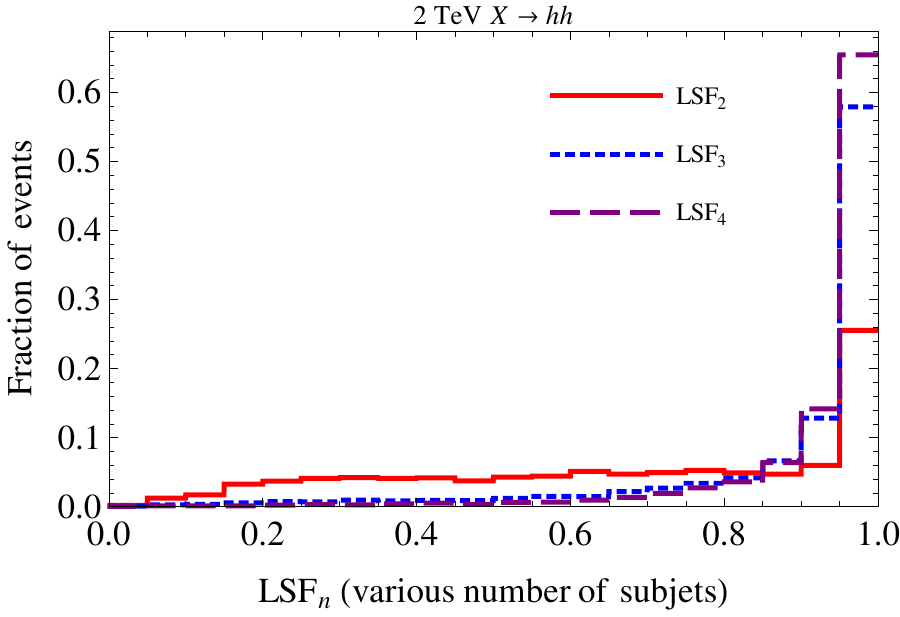}
                \vspace{-5mm}
                \label{fig:LSFn_hh}
                \caption{}
        \end{subfigure}
\hspace{5mm}
        \begin{subfigure}[b]{7.0cm}
\includegraphics[width=7.0cm]{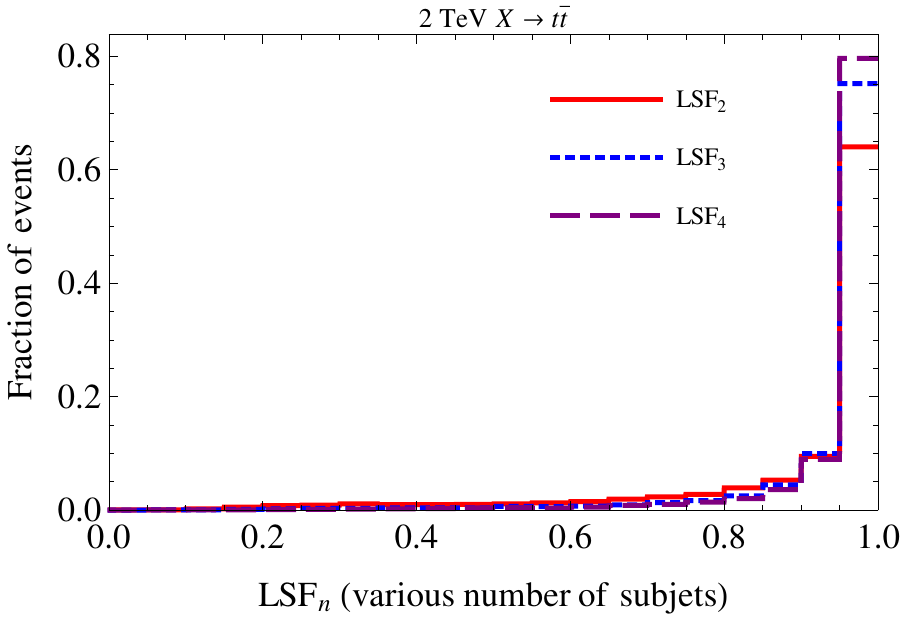}
                \vspace{-5mm}
                \label{fig:LSFn_tt}
                \caption{}
        \end{subfigure}

\caption{Distributions of LSF for the leading lepton as computed with two ($\text{LSF}_2$), three ($\text{LSF}_3$) or four ($\text{LSF}_4$) subjets, for various event samples.}
\label{fig:LSFn}
\end{figure}

Figure~\ref{fig:LSFn} shows distributions of LSF computed with different numbers of subjets (two, three and four) for various background and signal models. These confirm the behavior discussed above. Figure~\ref{fig:LSFn_bb} shows distributions for a $b \bar{b}$ background sample, illustrating how LSF shifts to higher values as the number of subjets increases. For the three-parton signal models, squark-neutralino and di-Higgs resonance, LSF shifts sharply downwards as the number of subjets is lowered from three to two. However, for the $t \bar{t}$ resonance model, LSF remains high for any number of subjets. These results confirm that optimal sensitivity is achieved when the number of subjets is matched to the number of partons from the boosted decay; however if more subjets are used then signal efficiency remains high, with the only penalty being a moderate increase in background. Therefore for searches designed to probe multiple types of models, such as a generic search for events with high $H_T$ and hard non-isolated leptons, the choice of $\text{LSF}_n$ variable should be optimized for the targeted model with the largest number of partons from the boosted decay.

%%%%%%%%%%%%%%%%%%%%%%%%%%%%%%%%%%%%%%%%%%%%%%%%%%%%%%%%%%%%%%%%%%%%%%%%%%%%%%%%
%%%%%%%%%%%%%%%%%%%%%%%%%%%%%%%%%%%%%%%%%%%%%%%%%%%%%%%%%%%%%%%%%%%%%%%%%%%%%%%%
%%%%%%%%%%%%%%%%%%%%%%%%%%%%%%%%%%%%%%%%%%%%%%%%%%%%%%%%%%%%%%%%%%%%%%%%%%%%%%%%

\section{Proof-of-principle search: RPV SUSY}
\label{sec:search}

As a concrete example of potential searches involving non-isolated leptons, we now consider a hypothetical search targeting the squark-neutralino model ($\tilde{q} \rightarrow q \tilde{\chi}, \tilde{\chi} \rightarrow\ell qq$) presented in section \ref{sec:motivation}. As shown in \cite{Graham:2014vya}, this model provides an example of new physics parameter space which could be discovered in the 8 TeV data with improved search strategies. This simulated search is not intended to be an optimized strategy for discovering this particular model, but rather a proof-of-principle demonstration of our proposed approach. We expect that this search can be improved with further work and that our basic proposal can be extended to a wider class of models.

As we can see from figure \ref{fig:squark-topology}, our signal events typically contain a large amount of hadronic activity, as well as two hard non-isolated leptons. We therefore place the following hard baseline kinematic cuts: 
\begin{itemize}
\item cluster hadronic jets (i.e.\ \emph{not} including leptons\footnote{As before, we still cluster leptons that fail the baseline cuts of $p_T > 20$ GeV, $|\eta| < 2.5$, but impose no isolation requirements on harder leptons.}) using the anti-$k_T$ algorithm with radius parameter $R_{\textrm{jet}} = 0.5$,
\item require at least four jets with $p_T > 150$ GeV and $|\eta| < 3.0$,
\item require at least two leptons with $p_T > 40$ GeV and $|\eta| < 2.5$,
\item require $H_T > 850$ GeV, where we define
\end{itemize}
\begin{equation}
H_T = \sum_j p_{T,j} + \sum_\ell p_{T,\ell}.
\end{equation}
These rather aggressive cuts are specifically designed to greatly reduce the SM backgrounds, while retaining a large signal efficiency. After the above kinematic cuts, we find that the SM background consists mainly of QCD multijet and $t \bar{t}$ events. However, due to the large SM cross-sections for these processes, enough background remains to necessitate other means of differentiating the signal events. We will consider various possible searches using the lepton isolation variables discussed in the previous section and compare their reach in the model parameter space.

To obtain the QCD background, we generated four-jet inclusive samples in MadGraph 5.1.5.7, then matched, showered and hadronized them using Pythia 8.1.85, and normalized the cross-section by multiplying by a global $k$-factor obtained by comparing measured~\cite{Aad:2011tqa} versus Monte Carlo inclusive multijet cross-sections subject to various cuts. There is likely significant error in this estimate of the QCD background cross-section, but we expect that this approximation is sufficient for the purposes of our demonstration, particularly since we find QCD to be a subdominant component of the background in the cases of interest. In a real experimental study, QCD backgrounds could be determined by data-driven techniques.

%do not expect that this represents the true QCD cross section for our search (as would be determined by a data-driven analysis or comparable technique in a true experimental study), but we expect that our estimate is a decent enough approximation for the purposes of our toy search.

 %In the former, the two hard leptons originate from decays of hadrons, while the latter catergory includes events with one or more leptons from $W$ bosons. However, in $t \bar{t}$ events,  

\begin{figure}[t!]
\centering

\vspace{5 mm}

\includegraphics[width=7.0cm]{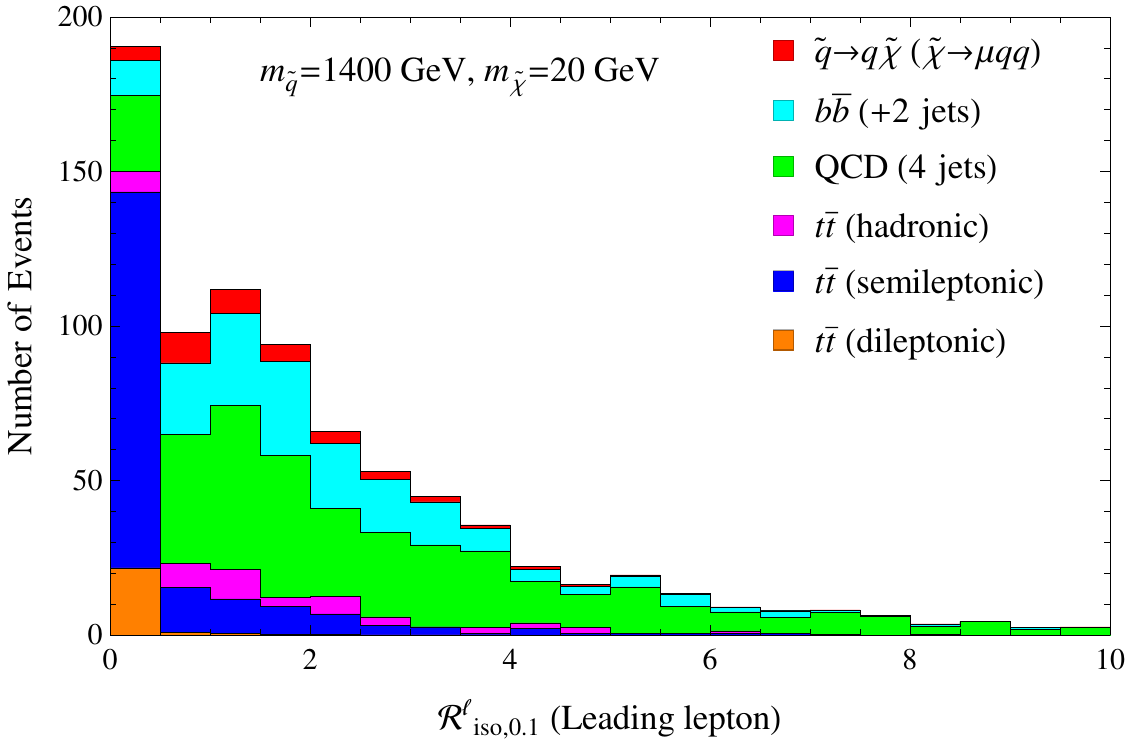}
\hspace{1.5mm}
\includegraphics[width=7.0cm]{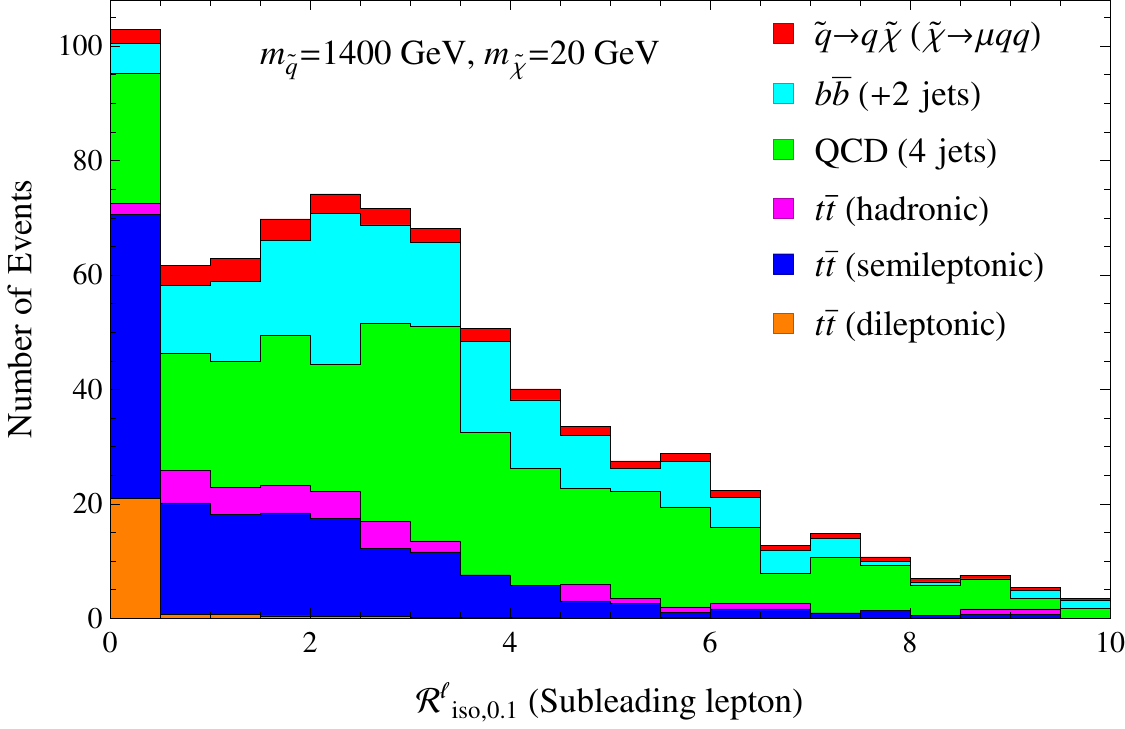}
\caption{Number of events as a function of $\reliso$ in a cone of $R_{\textrm{cone}} = 0.1$ for the hardest (left) and second hardest (right) leptons, for the $\tilde{q}\rightarrow q \tilde{\chi}$ model and relevant SM backgrounds. These events have passed the baseline cuts listed in this section and are normalized to correspond to the expected number of events in the 8 TeV 20 fb$^{-1}$ dataset. The ``$b \bar{b}$" category includes QCD multijet events generated with two $b$ quarks at the matrix element level. The signal parameters are chosen to be $m_{\tilde{q}} = 1400$ GeV, $m_{\tilde{\chi}} = 20$ GeV. The gluino mass is set to 2 TeV for all models, and only squark pair production is considered.}
\label{fig:search-RelIso}
\end{figure}

Figure \ref{fig:search-RelIso} shows the expected number of signal and background events in the 8 TeV LHC data as a function of the relative isolation $\reliso$ for the two hardest leptons, computed with a cone size of $R_\textrm{cone} = 0.1$ (compared with $R_\textrm{cone} = 0.3 - 0.4$ for standard LHC searches). The signal distribution is shown for the benchmark point of $m_{\tilde{q}} = 1400$ GeV and $m_{\tilde{\chi}} = 20$ GeV, with a production cross-section of 4.1 fb (obtained using NLL-fast \cite{Beenakker:2011fu} with gluino mass $m_{\tilde{g}}=2$ TeV). Both the signal and background events were simulated using the software listed in section \ref{sec:variables}, and the resulting distributions are normalized to correspond to the correct number of events in the full 20 fb$^{-1}$ dataset from the 8 TeV LHC. Note that the expected signal is within approximately an order of magnitude of the background for this benchmark point, as the hard cuts we required eliminate much of the SM backgrounds. However, after these cuts the relative isolation variable does not appear particularly effective in further separating signal and background. 

%\begin{itemize}
%\item $m_{\tilde{q}} = 1000$ GeV and $m_{\tilde{\chi}} = 100$ GeV, with a production cross-section of $48$ fb,
%\item $m_{\tilde{q}} = 1400$ GeV and $m_{\tilde{\chi}} = 100$ GeV, with a production cross-section of $4$ fb.
%\end{itemize}
\begin{figure}[t!]
\centering

\vspace{5 mm}

\includegraphics[width=7.0cm]{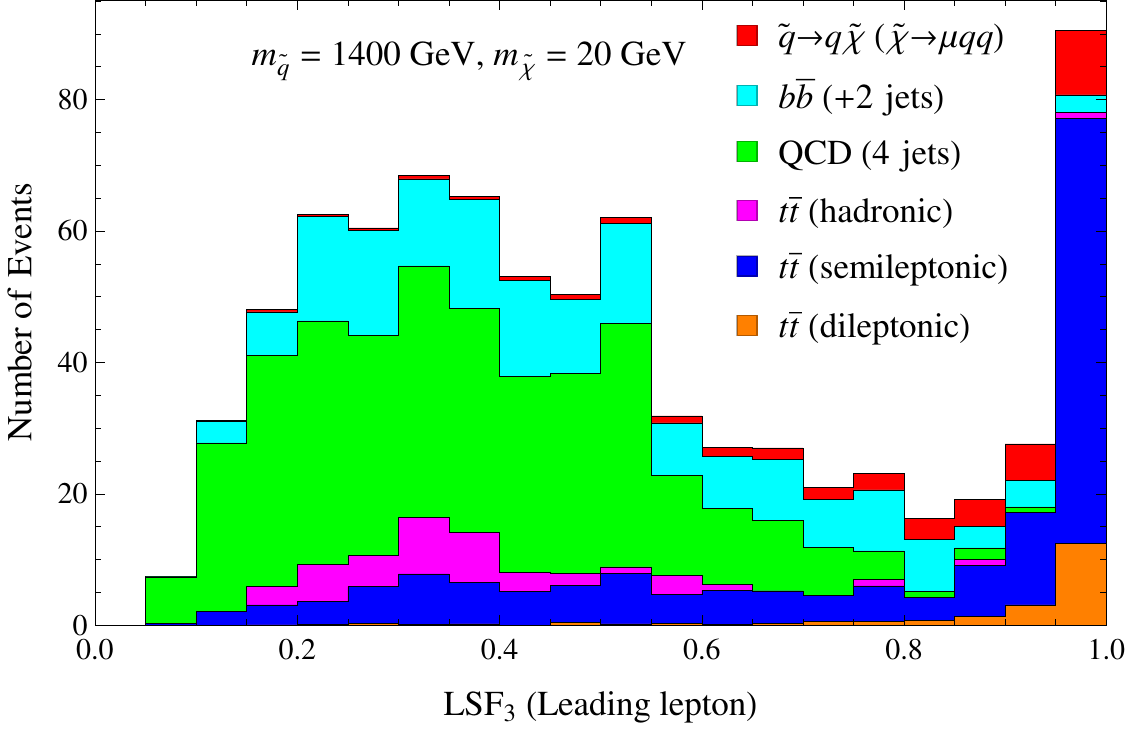}
\hspace{1.5mm}
\includegraphics[width=7.0cm]{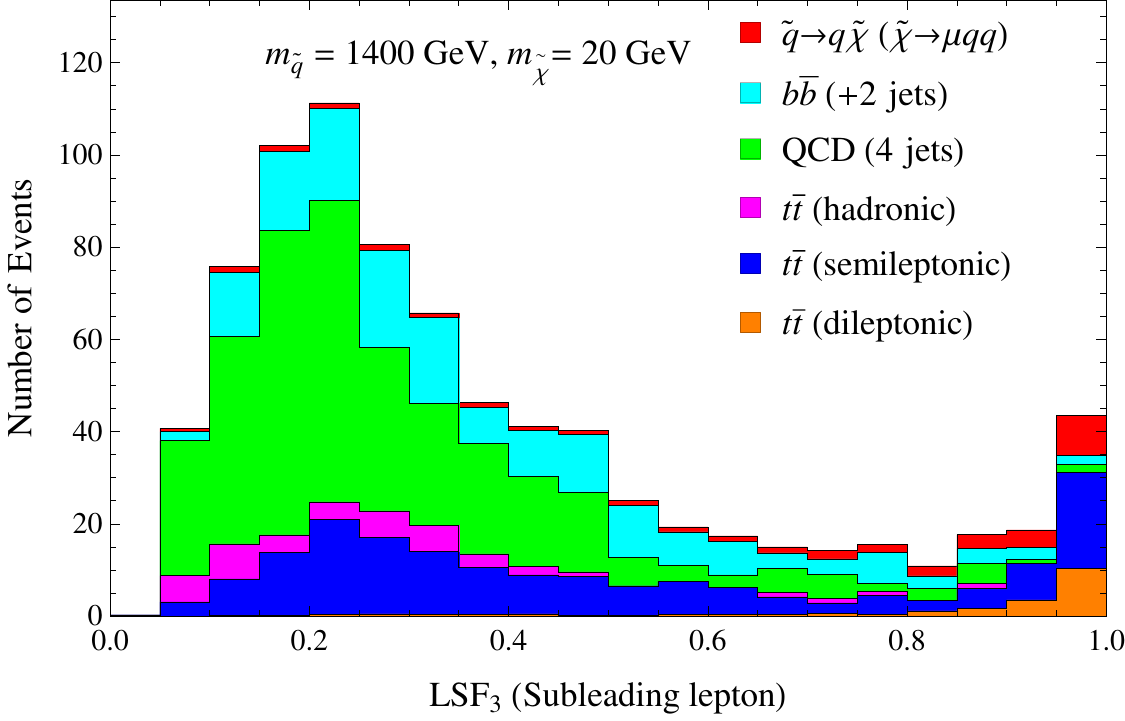}

\caption{Lepton subjet fraction ($\text{LSF}_3$) for the hardest (left) and second hardest (right) leptons, for the $\tilde{q}\rightarrow q \tilde{\chi}$ model and relevant SM backgrounds. These events have passed the baseline cuts listed in this section and are normalized to correspond to the expected number of events in the 8 TeV 20 fb$^{-1}$ dataset. The signal parameters are chosen to be $m_{\tilde{q}} = 1400$ GeV, $m_{\tilde{\chi}} = 20$ GeV. The gluino mass is set to 2 TeV for all models, and only squark pair production is considered.}
\label{fig:search-LepFraction}
\end{figure}

\begin{figure}[t!] 
\centering
\includegraphics[width=7.0cm]{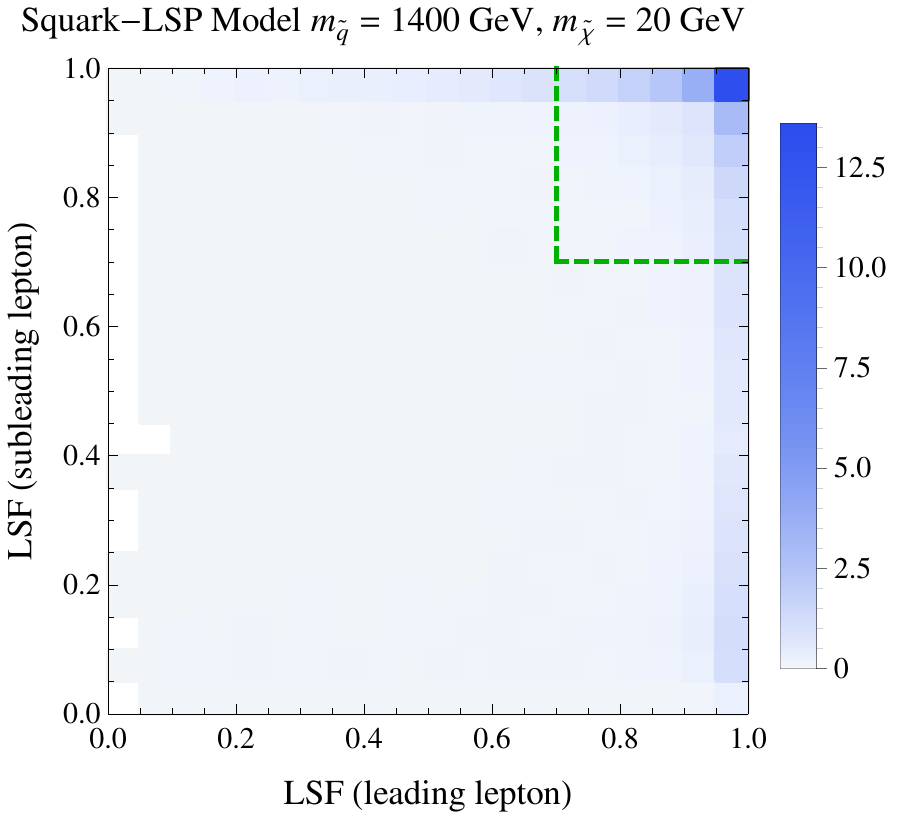}
\hspace{1.5mm}
\includegraphics[width=7.0cm]{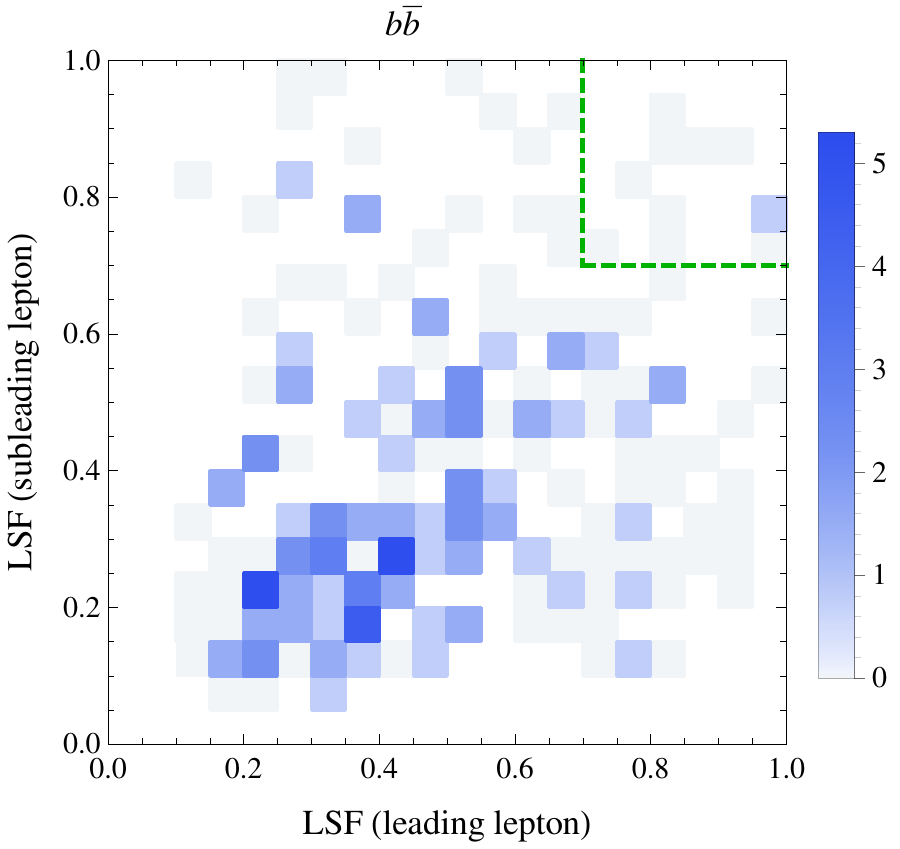}

\vspace{5 mm}

\includegraphics[width=7.0cm]{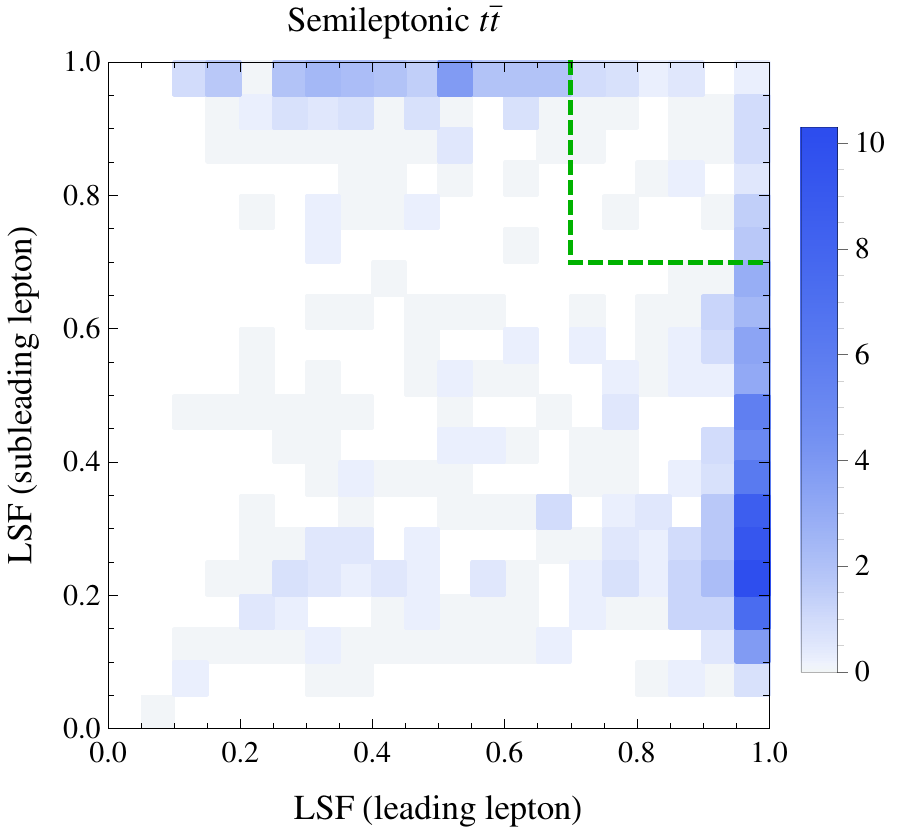}
\hspace{1.5mm}
\includegraphics[width=7.0cm]{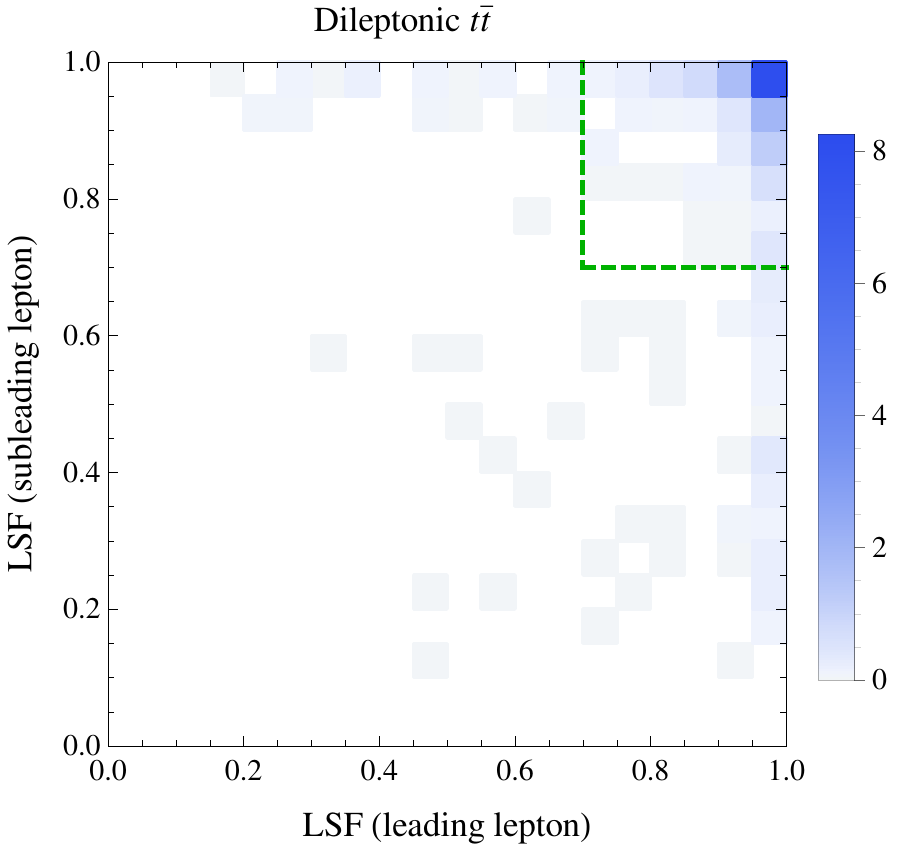}
\caption{2D histogram of $\text{LSF}_3$ of the leading and subleading leptons for various event samples after kinematic cuts. The green dashed box indicates a choice of signal region corresponding to the requirement $\text{LSF}_3 > 0.7$ for both leptons. The scales on the legends indicate the expected number of events in $20 \text{ fb}^{-1}$ of 8 TeV data. }
\label{fig:2DLSF}
\end{figure}
In figure~\ref{fig:search-LepFraction} we again show the expected signal and background counts, now as a function of the lepton subjet fraction (LSF) of the two leptons. We see that even after kinematic cuts LSF provides excellent discrimination between the signal and the QCD (including $b \bar{b}$) background. However, SM $t \bar{t}$ events can produce truly isolated leptons from $W$ bosons, which will have high LSF. Semileptonic $t \bar{t}$ events only produce one isolated lepton and can therefore be suppressed by requiring two leptons with high LSF. Dileptonic $t \bar{t}$ produces two isolated leptons but has low efficiency to pass the four jet kinematic cut we require. (Similarly, electroweak boson production with decays to leptons very rarely passes the kinematic cuts and is negligible for this study.) Figure~\ref{fig:2DLSF} shows the distribution of LSF of the two hardest leptons in each event for both the squark-neutralino signal and the relevant backgrounds. As we can see, LSF cuts on both leptons greatly reduce the background from events with leptons originating from jets, such that the dominant background becomes dileptonic $t \bar{t}$.

\begin{figure}[t!] 
\centering
\includegraphics[width=10.0cm]{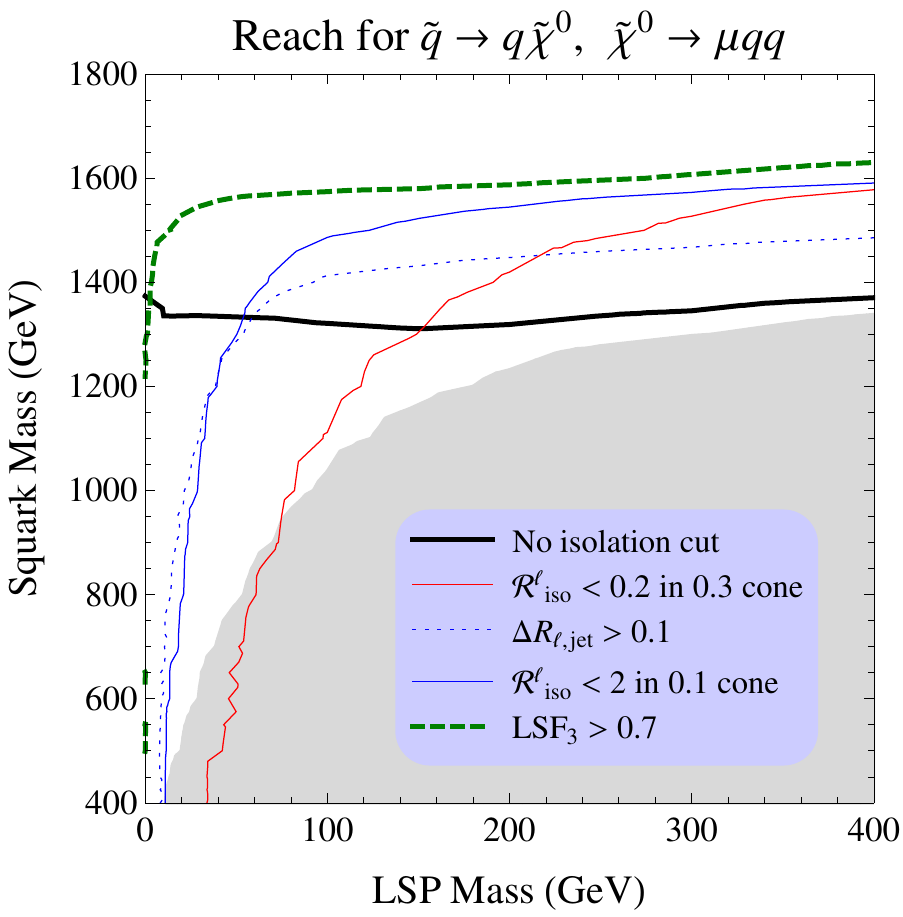}
\caption{Projected exclusion sensitivity of the 8 TeV LHC data at $95\%$ confidence level for the $\tilde{q}\rightarrow q \tilde{\chi}$ model, using the various search strategies proposed in this section. The shaded region corresponds to the exclusion limits already placed by current searches~\cite{Graham:2014vya}, while the various curves mark the exclusion reaches of hypothetical cut-and-count searches using our hard kinematic cuts and various choices of lepton isolation cuts (see discussion in text). The gluino mass is set to 2 TeV for all models, and only squark pair production is considered.}
\label{fig:scan}
\end{figure}

To get a complete picture of how various lepton isolation variables perform in probing this model, we plot the exclusion reach in the parameter space of squark mass $m_{\tilde{q}}$ and neutralino mass $m_{\tilde{\chi}}$ in figure~\ref{fig:scan}. The shaded region corresponds to the existing bounds on this model based on current LHC data, as claimed in~\cite{Graham:2014vya} from a reinterpretation of a CMS search in events with same-sign leptons, which required $\reliso < 0.15$  with $R_\textrm{cone} = 0.3$ for both signal leptons. The resulting bound on the squark mass rapidly erodes as the neutralino is made sufficiently light. This drop in sensitivity occurs because the light neutralinos become highly boosted, such that the leptons tend to not be isolated. 

The other curves show the reach for hypothetical cut-and-count searches using the above baseline kinematic cuts and various different lepton isolation criteria. To obtain these estimates we generated Monte Carlo signal samples for a grid of points in this plane and determined the expected signal and background counts for the various searches. The contours indicate the points in model space which are expected to be excluded at 95\% confidence level if only SM events are observed.

%, deeming a model point excluded if it predicted more than $2\sqrt{N_B} +3$ $\textbf{WHY?}$ signal events, where $N_B$ is the number of counts expected from background. 

The solid horizontal black curve in figure~\ref{fig:scan} corresponds to the bound achieved if no lepton isolation cut is applied to events. While this admits many QCD background events, it still performs better than the existing CMS same-sign lepton search, due to the hard kinematic cuts that were tailored for this particular signal. In particular, it is far superior for light (highly boosted) neutralinos where standard lepton isolation rejects most signal events. The solid red curve corresponds to a search with the baseline kinematic cuts and a ``standard'' isolation criterion $\reliso < 0.2$ with $R_\textrm{cone} = 0.3$ for both leptons. The addition of an isolation criterion improves the reach at high $m_{\tilde{\chi}}$ (the unboosted regime), but again performs much worse for even moderately boosted neutralinos. The solid blue curve corresponds to a more relaxed relative isolation cut, requiring $\reliso < 2$ with $R_\textrm{cone} = 0.1$. This modified search has greatly improved sensitivity, especially at low $m_{\tilde{\chi}}$, but here also the bound sharply drops off for sufficiently boosted neutralinos. Similar bounds are achieved by a cut on $\Delta R$ between a lepton and the nearest jet of $\Delta R_{\ell,j} > 0.1$ (dotted blue curve). Finally, the dashed green curve shows the bound acheived with a cut of lepton subjet fraction (LSF) $> 0.7$ for both leptons. Here the reach in the $m_{\tilde{q}}$ (corresponding to total cross-section) is high and roughly constant with $m_{\tilde{\chi}}$ down to extremely large boost. This reflects the fact that LSF allows for considerable background rejection while retaining signal efficiency for a large range of neutralino boosts. 

While this hypothetical search demonstrates the utility of the LSF variable, there are several potential improvements that could further enhance the signal reach. One general technique is to use additional jet substructure variables to specifically select the three-axis (two jet plus one lepton) topology that is characteristic of signal, allowing further rejection of leptonic tops and other backgrounds. For example, the $N$-subjettiness variables $\tau_{31}$ and $\tau_{32}$~\cite{Thaler:2010tr} could be used to select signals with moderately boosted neutralinos (though potentially at the expense of efficiency in the unboosted case). For this particular RPV model, it would also be useful to consider the invariant masses of the fat jets containing the leptons, as these should approximately reconstruct the neutralino mass (with increased accuracy achieved with the use of jet grooming techniques, e.g.~\cite{Butterworth:2008iy,Krohn:2009th,Ellis:2009su,Ellis:2009me,Krohn:2013lba,Larkoski:2014wba,Berta:2014eza,Cacciari:2014jta,Cacciari:2014gra,Bertolini:2014bba}). Using information about the distribution of events in this variable would allow for better constraints compared to the simple cut-and-count approach described above. One could also imagine potential refinements to the LSF variable itself; one interesting avenue for study would be the discrimination efficiency of LSF with the use of alternative recombination schemes, such as the ``winner-take-all'' scheme presented in \cite{Bertolini:2013iqa,Larkoski:2014uqa,Larkoski:2014bia}. Finally, in an actual experimental study the QCD background rates would not be estimated purely from Monte Carlo but instead be obtained using data-driven techniques. For example, cuts on $\tau_{31}$ and $\tau_{32}$ could be used to define signal-depleted control regions enriched in leptons from heavy quark jets. A veto on $b$-tagging (using e.g.\ a tight operating point) could be used to remove semileptonic top decays with negligible impact on signal efficiency and, at the same time, define a control region enriched in signal-like leptons (from $W$ boson decays) and very boosted $b$-jets, both of which could then be used for a data-driven calibration of LSF distributions.

%Figure \ref{fig:scan} shows the experimental sensitivity possible with the search strategies presented here, as compared to the current exclusion limits placed by searches with lepton isolation requirements. This exclusion plot was generated by PRASHANT PLEASE INSERT INFORMATION ON THE SCAN HERE. 

\section{Conclusions}
\label{sec:conclusions}

The upcoming 13 TeV run of the LHC presents an opportunity to re-evaluate approaches to searching for new physics. Given the current lack of experimental evidence for physics beyond the SM thus far, it is important to use every available tool to ensure that no potentially observable signals are missed. In this work, we have demonstrated the utility of non-isolated leptons in searches for boosted particles. This useful signature generically arises in decay chains with a hierarchy of masses and can provide sensitivity to a variety of models.

Here we have considered a basic approach to utilizing non-isolated leptons, combining strong cuts on hadronic activity and substructure analyses with no built-in size parameter. This strategy allows for the use of new isolation variables, lepton subjet fraction and lepton mass drop, which efficiently discriminate various signals from the dominant SM backgrounds. As we have shown, with this conceptually simple approach there is potential to discover new physics even using only the 8 TeV data.

While we have provided a proof-of-principle of this approach to using non-isolated leptons, further investigation is warranted to design searches with optimized discovery reach. Such searches are likely to benefit from the use of additional substructure-based variables, such as jet mass and $N$-subjettiness discussed above. It may also be possible to improve on LSF as defined in this work with related but more sophisticated substructure variables. Also of interest are techniques to obtain data-driven estimates of the backgrounds to searches using non-isolated leptons.

%, which will require understanding of how LSF distributions vary with event topology.  

%We expect the LSF distributions of individual objects such as QCD jets, boosted tops etc. to be independent of the other activity in an event, so that such techniques should be viable.  

%This work is only a first step in the use of non-isolated leptons. While LSF and LMD provide a robust means of distinguishing signal from background, we expect that our approach can be further improved with future work and the development of more effective discriminants. A survey of more sophisticated substructure analyses of non-isolated leptons would be a useful direction for future research.

We encourage both CMS and ATLAS to not only revisit past data with an eye towards utilizing non-isolated leptons, but also to develop new triggers based on the combination of hard hadronic activity with moderately hard leptons, without the requirement for standard lepton isolation. Reliance solely on hadronic triggers with very high $p_T$ requirements can significantly reduce signal efficiency for many models, limiting the experimental reach possible with the LHC. Provided that new physics events are adequately triggered on, non-isolated leptons can be a promising tool for discovering them within the data.  

%A full study of the use of non-isolated leptons can further develop and capitalize on this promising means of discovering new physics.

\acknowledgments{We are grateful to Ian Anderson, Marc Osherson, Jesse Thaler and Itay Yavin for helpful discussions. We also thank Andrew Larkoski for helpful comments on the draft. CB was supported under NSF grant PHY-1315155, and is also grateful for support from the Maryland Center of Fundamental Physics. The research of CB was supported in part by the Perimeter Institute for Theoretical Physics. Research at Perimeter Institute is supported by the Government of Canada through Industry Canada and by the Province of Ontario through the Ministry of Economic Development and Innovation. PM and YX were supported by NSF grant PHY-1404302. AS was supported by the Clare Boothe Luce Program of the Henry Luce Foundation with award number 7815. PS acknowledges the support of the Maryland Center for Fundamental Physics, and NSF grants PHY-1315155 and PHY-1214000. MTW was supported by the John Templeton Foundation, NSF grant PHY-1214000 and DOE grant DE-SC0010025. Finally, we thank the University of Maryland, College Park Physics Department for computing time on the UMD HEP T3 Computing Cluster and Stanford University for the use of the FarmShare research computing environment.

\bibliography{leptojet_reformatted_refs}
\bibliographystyle{JHEP}

\end{document}